\documentclass[prl,superscriptaddress,floatfix,nofootinbib,twocolumn,aps,10pt]{revtex4-2}
\usepackage{graphicx,amssymb,amsmath,bm,dcolumn,xcolor,mathtools}
\usepackage[caption=false]{subfig}
\usepackage[colorlinks,linkcolor=red,citecolor=blue,urlcolor=blue]{hyperref}
\usepackage{url}
\usepackage{upgreek}
\usepackage{threeparttable}
\usepackage{comment}
\usepackage{xfrac}
\usepackage{appendix}
\usepackage{enumitem}
\usepackage[normalem]{ulem}
\usepackage{soul}
\usepackage{multirow}
\newcounter{parnum}
\begin{document}

\title{Nanofabricated torsion pendulums for tabletop gravity experiments}

\author{J. Manley}
\affiliation{National Institute of Standards and Technology, Gaithersburg, MD 20899, USA}

\author{C. A. Condos}
\affiliation{Wyant College of Optical Sciences, University of Arizona, Tucson, AZ 85721, USA}

\author{Z. Fegley}
\affiliation{National Institute of Standards and Technology, Gaithersburg, MD 20899, USA}
\affiliation{Department of Physics, University of Maryland, College Park, MD 20742, USA}

\author{G. Premawardhana}
\affiliation{Joint Quantum Institute/Joint Center for Quantum Information and Computer Science, University of Maryland-NIST, College Park, MD 20742, USA}

\author{T. Bsaibes}
\affiliation{National Institute of Standards and Technology, Gaithersburg, MD 20899, USA}
\affiliation{Department of Physics, University of Maryland, College Park, MD 20742, USA}

\author{J. M. Taylor}
\affiliation{Joint Quantum Institute/Joint Center for Quantum Information and Computer Science, University of Maryland-NIST, College Park, MD 20742, USA}

\author{D. J. Wilson}
\affiliation{Wyant College of Optical Sciences, University of Arizona, Tucson, AZ 85721, USA}

\author{J. R. Pratt}
\affiliation{National Institute of Standards and Technology, Gaithersburg, MD 20899, USA}

\begin{abstract}
Measurement of mutual gravitation on laboratory scales is an outstanding challenge and a prerequisite to probing theories of quantum gravity. A leading technology in tabletop gravity experiments is the torsion balance, with limitations due to thermal decoherence. Recent demonstrations of lithographically defined suspensions in thin-film silicon nitride with macroscale test masses suggest a path forward, as torsion pendulums dominated by gravitational stiffness may achieve higher mechanical quality factors through dilution of material losses. Here we demonstrate a 250~$\upmu$m~$\times$~5~mm~$\times$~$1.8$~$\upmu$m torsion fiber supporting 87 grams and forming a Cavendish-style torsion pendulum with tungsten test masses that---to our knowledge---is the largest thin-film silicon-nitride-based oscillator to date. Torsion pendulums with thin-film, nanofabricated suspensions provide a test bed for near-term tabletop experiments probing classical and quantum gravitational interaction between oscillators.   
\end{abstract}

\maketitle

\section{Introduction}
Torsion balances have long been central to laboratory gravity experiments~\cite{gillies1993torsion}, from the Cavendish experiment~\cite{cavendish1798xxi} to modern tests of the Equivalence Principle and the gravitational inverse square law~\cite{adelberger2003tests,adelberger2009torsion}. Their combination of low stiffness and large test masses yields sensitivity to the minute gravitational fields of local source masses. As a result, proposals to test quantum gravity commonly suggest experiments based on torsion balances~\cite{westphal2021measurement,komori2020attonewton,agafonova20241,lami2024testing,matsumura2020gravity,yan2025first,kafri2013noise,al2018optomechanical}. And while there has been a recent surge in interest to develop quantum gravity experiments~\cite{miao2020quantum,miki2024quantum,datta2021signatures,krisnanda2020observable,miki2024feasible,beyer2025one,tang2025optimal,kryhin2025distinguishable,zhong2025distinguishing,carney2021using}, there has not yet been an experiment capable of detecting the two-way gravitational interaction between laboratory objects, a prerequisite to the observation of gravity-induced quantum entanglement~\cite{tang2025cavity,westphal2021measurement}.

\begin{figure*}
    \centering
    \includegraphics[width=1.98\columnwidth,trim= 0in 0in 0in 1in]{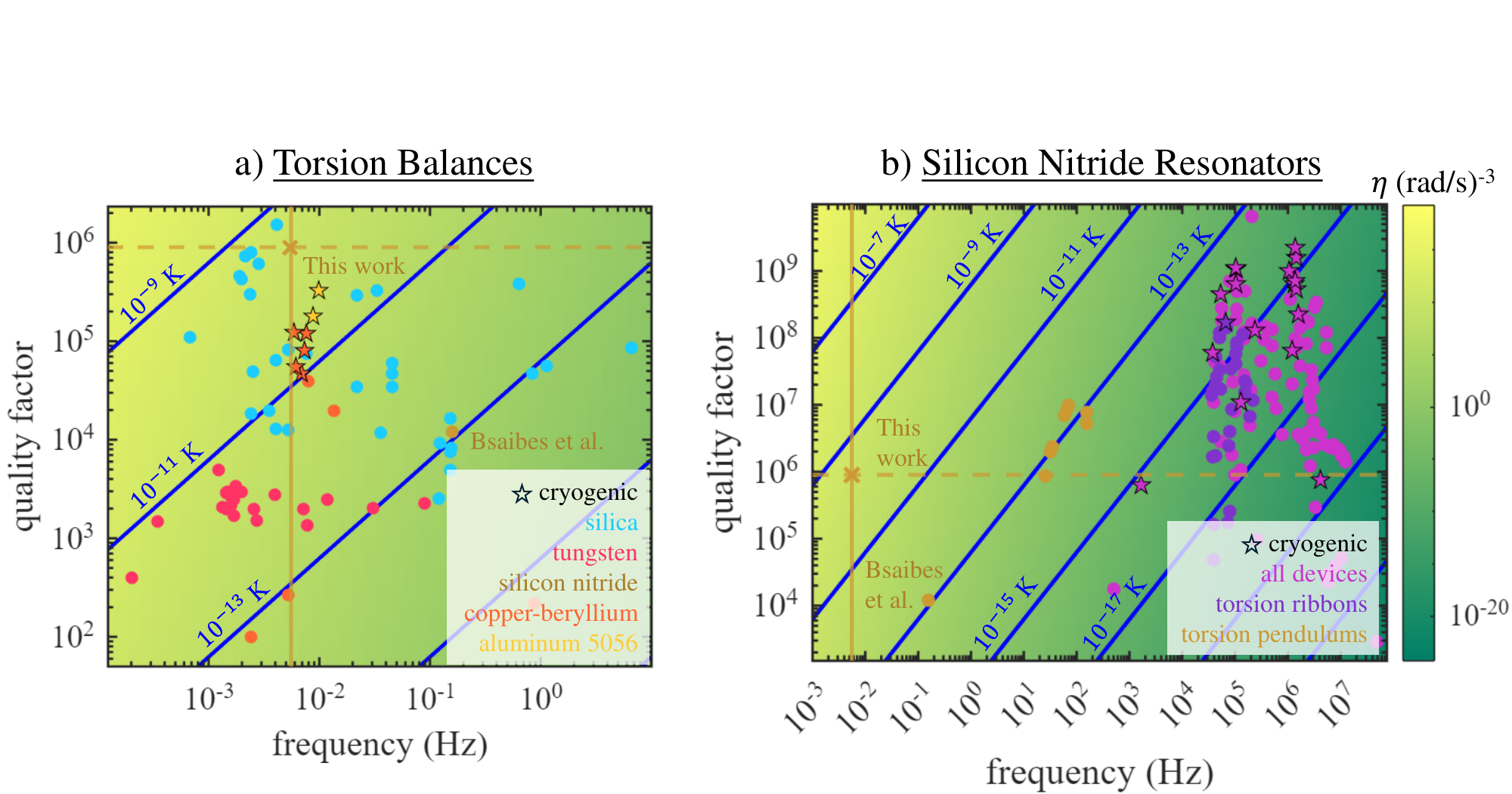}
    \caption{Compilations of torsion balances and thin-film silicon nitride resonators. The plots are underlaid by the mechanical design figure of merit $\eta=Q/\omega_0^3$ and include contours indicating the cryogenic cooling necessary for gravitational entanglement to overcome thermal decoherence, assuming spherical tungsten test masses. The torsion pendulum presented in this work is marked `x', with measured frequency (theoretical $Q$) marked by a brown, solid vertical (dashed horizontal) line. \textbf{a)} Compilation of torsion balances with quartz/silica~\cite{downsbrough1937damping,zhao2023experimental,li2018measurements,li2014g,yang2009direct,ross2025probing,liu2025amplitude,shaw2022torsion,westphal2021measurement,hagedorn2006quality,cavalleri2009new,agafonova20241,fu2025study,catano2020high,su2024influence} and metallic~\cite{hu2000amplitude,chen2009nanonewton,li2014g,yang2009direct,quinn1995stress,richman1999preliminary,newman2014measurement,newman1999determining,bantel2000high,fleischer2022cryogenic,ubhi2025demonstration,tu2010performance,tu2009electrostatic,yang2012torsion,lin2025new,xu2025measuring,Schlamminger2008,gundlach1997short,tan2016new,Kapner2007,yang2012test,rajalakshmi2008torsion,abercrombie2016development,carbone2005characterization,hueller2002torsion,ciani2017new,cavalleri2009new,bai2015improving,xu2022measuring,dong2023coupling} suspensions. \textbf{b)} Compilation of Si$_3$N$_4$ thin-film resonators~\cite{liu2021gravitational,sadeghi2020frequency,wilson2009cavity,thompson2008strong,villanueva2014evidence,reinhardt2016ultralow,chowdhury2025optomechanical,tsaturyan2017ultracoherent,ghadimi2018elastic,chen2020entanglement,xia2023entanglement,thomas2021entanglement,zwickl2008high,zhang2012synchronization,anetsberger2009near,hyatt2025ultrahigh,wilson2015measurement,sudhir2017quantum,bereyhi2022hierarchical,gisler2022soft,fischer2016optical,yuan2015silicon,norte2016mechanical,chowdhury2023membrane,cupertino2024centimeter,hodges2023characterization} including torsion ribbons~\cite{shin2025active,pratt2023nanoscale,hyatt2025ultrahigh}. Brown points indicate mass-loaded torsion pendulums~\cite{bsaibes2025lithographically,manley2024microscale,condos2025ultralow}}
    \label{fig:compilation}
\end{figure*}

Thermal decoherence poses a stringent challenge to next generation gravity experiments with torsion balances. Material loss in metallic torsion fibers typically limits the mechanical quality factor to $Q\lesssim 10^4$ at room temperature (see Fig.~\ref{fig:compilation}a), resulting in thermal decoherence rates $\Gamma_\text{th} \approx k_\text{B} T /\left(\hbar Q\right) \gtrsim 2\pi\times 1$ GHz. In contrast, realistic gravitational interaction rates between oscillators would be at most $\lesssim 0.1$ mHz~\cite{kafri2014classical}, such that gravity-mediated phonon exchange between the oscillators would be overwhelmed by exchange with their independent thermal baths. Decoherence can be reduced through cryogenic cooling, which has the potential to lower dissipation~\cite{newman1999determining,bantel2000high} in addition to simply lowering the thermal occupation. Alternatively, torsion fiber materials with less intrinsic loss are being pursued, with silica suspensions achieving quality factors exceeding $10^6$~\cite{liu2025amplitude}.

Dissipation dilution provides a strategy to meet strict coherence requirements that can exceed the reach of materials engineering and cryogenics alone. In this approach, the loss of a dissipative spring is diluted by introducing an additional, effectively lossless restoring force~\cite{fedorov2019generalized}. In this context it may be advantageous to use a torsion pendulum based on a bifilar suspension~\cite{heyl1930redetermination}, where twists couple to vertical displacement, such that the weight of the test masses provides a lossless torsional stiffness. The same effect can be realized with a single ribbon-like suspension~\cite{quinn1997novel} whose width greatly exceeds the thickness ($w\gg h$). The quality factor is enhanced relative to the intrinsic value $Q_\text{int}$ in proportion to the square of the aspect ratio, $Q\propto Q_\text{int} w^2/h^2$, analogous to strain-induced dissipation dilution in nanomechanical torsion resonators~\cite{pratt2023nanoscale}.

Nanoscale thin films naturally enable high-aspect-ratio suspensions for diluting material losses. In particular, high-$Q$ nanomechanical resonators are regularly realized in thin-film silicon nitride (Si$_3$N$_4$), spanning a broad range of frequencies from 10 Hz to 10 MHz~\cite{condos2025ultralow,pratt2023nanoscale,wilson2012cavity,aspelmeyer2014cavity} (see Fig.~\ref{fig:compilation}b). The scope of Si$_3$N$_4$-based resonators continues to expand, with recent demonstration of a sub-Hz torsion pendulum~\cite{bsaibes2025lithographically}. Standard nanofabrication processes yield lithographically defined geometries with nanoscale tolerances that can span a wide range of sizes, and a rapid production rate provided by wafer-scale parallel processing permits iterative testing and design. With the additional prospects of mode shape engineering to reduce clamping loss and stress-induced dissipation dilution for ultra-coherence~\cite{tsaturyan2017ultracoherent,ghadimi2018elastic}, high-stress Si$_3$N$_4$ has an established history with experiments in quantum sensing~\cite{chen2020entanglement,xia2023entanglement,sudhir2017quantum,sudhir2017appearance}, including demonstration of entanglement between mechanical objects~\cite{thomas2021entanglement}.

Here we explore nanofabricated torsion suspensions as a platform for tabletop gravity experiments, highlighting the challenge to probing both classical and quantum gravitational interactions between two torsion oscillators.  As a proof of principle, we demonstrate an 87 gram torsion pendulum with a $1.8$~$\upmu$m thick Si$_3$N$_4$ suspension---to our knowledge the largest thin-film Si$_3$N$_4$ oscillator to date (Fig.~\ref{fig:SiN_TB})---accessing a regime in which gravitational dissipation dilution becomes possible. Our result represents an important step toward ultracoherent, mHz-frequency, kg-scale torsion oscillators for next-generation gravity experiments.

\section{Nanofabricated torsion pendulums}
Bsaibes et al.~\cite{bsaibes2025lithographically} recently demonstrated a torsion pendulum with a thin-film Si$_3$N$_4$ suspension and a macroscale test mass. The device was etched from a silicon (Si) substrate coated with a 1.8~$\upmu$m thick layer of low stress Si$_3$N$_4$, where the 25~$\upmu$m~$\times$~25~mm~$\times$~$1.8$~$\upmu$m (width~$w$~$\times$~length~$l$~$\times$~thickness~$h$) suspension ribbon was released during the Si wet etch. The suspension supports a 37 mg  Si mass, which is expected to produce a mean stress 8 MPa in the ribbon. The fundamental torsion mode has a measured resonance frequency of $\omega_0=2\pi \times 160$ mHz and an intrinsic quality factor $Q_\text{int}\approx 1.2\times 10^4$, and demonstrated thermalization to 300~K. 

\begin{figure*}[t]
    \centering
    \includegraphics[width=1.98\columnwidth,trim= 0in 0in 0in 0in]{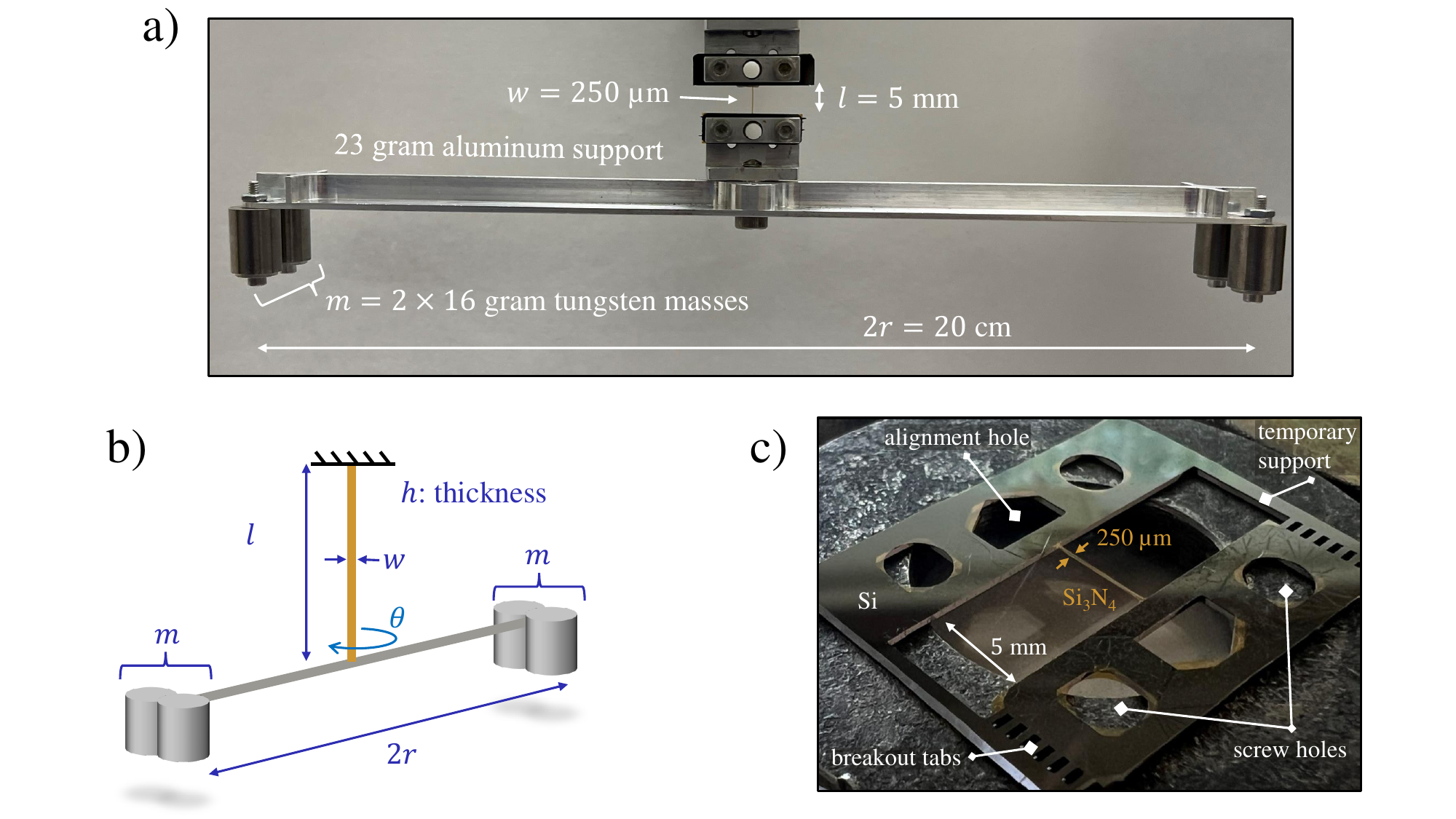}
    \caption{Macroscopic torsion pendulum with a nanofabricated suspension. \textbf{a)} A 5.6 mHz (180 s period) torsion pendulum formed by suspending an 87 gram mass from a 1.8~$\upmu$m thick Si$_3$N$_4$ ribbon suspension.  Cylindrical test masses are used for convenience; spherical masses are preferable for the gravitational coupling experiment described in the main text. \textbf{b)} Equivalent schematic, defining variables and design parameters. \textbf{c)} Image of the nanofabricated chip containing the suspension. The Si$_3$N$_4$ ribbon spans a window etched from a Si chip, with alignment and screw holes for clamping and temporary supports to protect the suspension during fabrication and mounting. The test mass is released by severing the breakout tabs and removing the supports.}
    \label{fig:SiN_TB}
\end{figure*}

To enable tabletop gravity experiments, it is necessary to significantly increase the size and density of the test mass. To this end, we have developed a higher-aspect-ratio ribbon and affixed large test masses as depicted in Fig.~\ref{fig:SiN_TB}, where an aluminum bar (23 grams) and four cylindrical tungsten masses (16 grams each) are supported by a 250~$\upmu$m~$\times$~5~mm~$\times$~$1.8$~$\upmu$m Si$_3$N$_4$ suspension. The torsion mode has an oscillation frequency of $5.6$ mHz (180 s oscillation period). With a total mass of 87 grams, the estimated mean stress of the suspension is 1.9 GPa. The test masses are located 10 cm from the torsion axis, such that the moment of inertia is calculated to be $7\times 10^{-4}$ kg m$^2$. The apparatus has been assembled and characterized in air, precluding a faithful measurement of the quality factor. In theory, the device would have a gravitational dilution factor $Q/Q_\text{int}= 74$ (see Appendix~\ref{app:pendulumModels}), reaching a total quality factor of $Q\approx 10^6$.

\section{Gravitational Coupling Experiments}\label{sec:GravExp}
Nanofabricated, thin-film suspensions are motivated by thermal noise and coherence requirements for measuring the dynamical gravitational coupling between torsion pendulums. In this section we discuss the potential for such experiments. We envision two identical torsion pendulums similar to that shown in Fig.~\ref{fig:SiN_TB}, each with a pair of tungsten spheres aligned in close proximity to produce a gravitational interaction, as depicted in Fig.~\ref{fig:gravityExpDesign}a.

\subsection{Classical gravity experiment} \label{sec:classGravExp}
We consider an experiment in which the mutual gravitation of two mechanical oscillators is probed by monitoring the frequency splitting between the differential and common modes~\cite{kafri2014classical}. For small oscillation amplitudes, the gravitation between the masses manifests as an additional spring that couples the masses. The coupled oscillator system has two mechanical modes (see Appendix~\ref{sec:modesplitting} for details), a breathing mode (differential motion) and center-of-mass mode (common motion) that are separated in frequency by 
\begin{equation} \label{eq:modesplitting}
    \Delta \omega \approx \frac{2Gm}{\omega_0 d^3}.
\end{equation}

Sensitivity would be limited by the imprecision of the mode frequency estimates. Thermal fluctuations limit the resolution to $\sigma_{\Delta\omega }=\sqrt{\omega_0/ Q t_\text{exp}}$ over an averaging time of $t_\text{exp}$. To achieve this resolution, readout noise of each oscillator position must be reduced such that thermal motion is resolved over the bandwidth $t_\text{exp}^{-1}$ (see Appendix~\ref{sec:sensitivityNoise} for details).

\begin{figure*}[t]
    \centering
    \includegraphics[width=1.98\columnwidth,trim= 0in 2.5in 0in 0in]{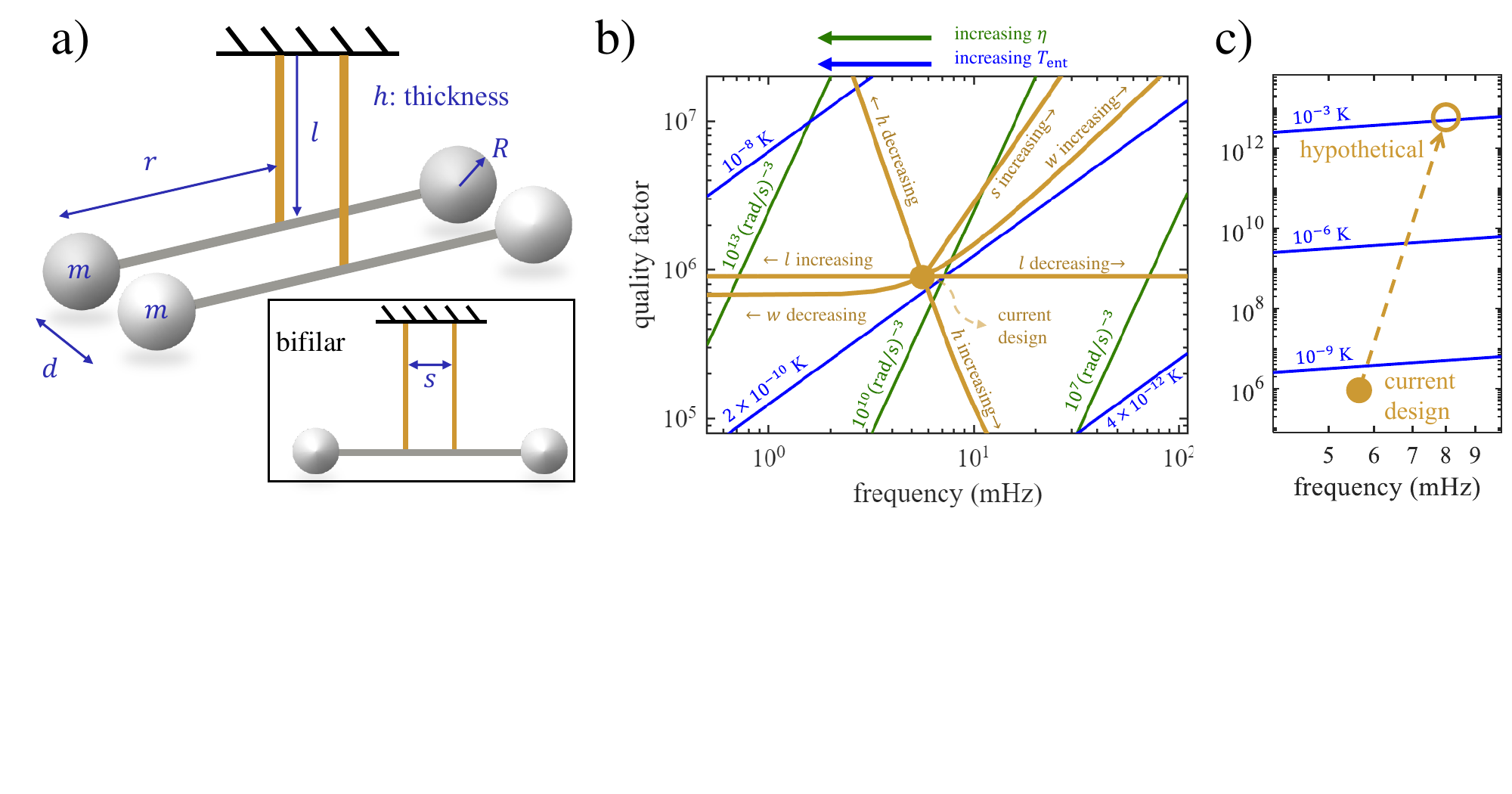}
    \caption{Design of a gravitational coupling experiment between two torsion pendulums. \textbf{a)} General configuration. \textbf{Inset:} Illustration of bifilar suspension. \textbf{b)} Visualization of design considerations, where brown curves indicate scaling of the pendulum quality factor and resonance frequency with suspension parameters, thickness $h$, length $l$, width $w$, and separation $s$, relative to the device in Fig.~\ref{fig:SiN_TB}. The suspension is assumed monofilar except for along the curve of increasing $s$. Blue (green) lines indicate contours of constant $T_\text{ent}$ ($\eta$), where in both cases the optimal designs favor high $Q$ and low $\omega_0$. \textbf{c)} Hypothetical improvement for the design described in Section~\ref{sec:gravEnt}. } 
    \label{fig:gravityExpDesign}
\end{figure*}

Optimal torsion balance design for a two-way gravity experiment would target low resonance frequency and high mechanical quality factor. We adopt a figure of merit (to be maximized)
\begin{equation}
    \eta\equiv Q/\omega_0^3
\end{equation}
to assess the performance of a torsion pendulum design. The origin of this figure of merit can be traced to the minimum measurement time needed to achieve a prescribed signal-to-noise ratio (defined here as $\Delta \omega/\sigma_{\omega_0}$) scaling as $t_\text{exp}\propto \eta^{-1}$ (see Appendix~\ref{sec:sensitivityNoise}). This scaling is visualized in Figure~\ref{fig:gravityExpDesign}b. In terms of suspension design parameters, this scaling translates to $\eta\propto w^{-1} h^{-2} l^{3/2}$ for a monofilar suspension and $\eta\propto s^{-1} h^{-2} l^{3/2}$ for a bifilar suspension with fibers separated by $s$ (see Appendix~\ref{app:pendulumModels} for the mechanical models used to derive this scaling), highlighting the advantage of thin films.

Measurement of the mutual gravitational dynamics between laboratory objects is well within reach for the prototype pendulum presented in Fig.~\ref{fig:SiN_TB}, with a frequency of $5.6$ mHz and a theoretical $Q\approx 10^6$. Consider an experiment positioning two identical pendulums with these parameters in close proximity to another, as depicted in Fig.~\ref{fig:gravityExpDesign}. Assuming replacement of the two cylindrical masses on each end with an equivalent spherical tungsten mass (a 32 g sphere with radius $7.3$ mm), and maintaining a large surface separation of $15$ mm ($d\approx 30$ mm) between the two pendulums to allow ample space for electrostatic shielding, a modesplitting of 0.7 $\upmu$Hz would be achieved, which can be detected with $2\sigma$ significance in just over two hours of measurement time (see Appendix~\ref{sec:sensitivityNoise} for details). This calculation assumes sensitivity limited by the pendulums' thermal motion, requiring readout sensitivity $\lesssim 30$ nrad/$\sqrt{\rm Hz}$ at the $5.6$ mHz resonance frequency, which is readily achievable with modern optical autocollimators~\cite{arp2013reference}.

\subsection{ Gravitational entanglement experiment} \label{sec:gravEnt}
In a quantum experiment to measure gravitational entanglement between two torsion pendulums, the mode frequency splitting (Eq.~\ref{eq:modesplitting}) again represents the observable effect, setting the scale for the two-mode squeezing of the mechanical modes of the coupled oscillator system~\cite{krisnanda2020observable,miki2024feasible,kafri2014classical,datta2021signatures}. Several proposals have considered optomechanical experiments in the quantum-noise-limited or backaction-limited regime~\cite{miao2020quantum,miki2024quantum,al2018optomechanical,datta2021signatures,liu2021gravitational}, where it has been shown that the minimum measurement time has the same dependence on mechanical properties as above, $t_\text{exp}\propto \eta^{-1}$~\cite{miao2020quantum}.

An entanglement experiment will additionally require that the gravitational interaction rate exceed the thermal decoherence rate, $\Delta \omega \gtrsim \Gamma_\text{th}$~\cite{kafri2014classical,miao2020quantum,miki2024quantum}, for entanglement to persist despite coupling to a thermal bath. This requirement imposes a strict bound on the maximum allowed temperature. Assuming tungsten test masses $\rho\approx 2\times10^4$ kg m$^{-3}$, the bound translates to
\begin{equation} \label{eq:temperatureRequirement}
    T_\text{ent}\lesssim  \left(\frac{Q}{1}\right) \left(\frac{1 \text{ Hz}}{\omega_0/2\pi}\right) 1.6 \times 10^{-18} \text{ K}
\end{equation}
From the models for $Q$ and $\omega_0$ in Appendix~\ref{app:pendulumModels}, the scaling with suspension design parameters can be shown to be $T_\text{ent}\propto w h^{-2} l^{1/2}$ ($T_\text{ent}\propto s h^{-2} l^{1/2}$ for bifilar). 
 
Equation~\ref{eq:temperatureRequirement} highlights the need for exceptionally low dissipation and therefore, an aggressive torsion balance design to enable entanglement at experimentally accessible (mK) temperatures.  We propose such a design in Fig.~\ref{fig:gravityExpDesign} based on bifilar thin-film suspensions (technical challenges are discussed in Sec.~\ref{sec:towardTabletop}).  Specifically, as shown in Fig.~\ref{fig:gravityExpDesign}a, we consider a pair of bifilar pendulums with suspension length $l=1$ m, thickness $h=50$ nm, widths $w=1$ mm, and ribbon separation $s=1$ cm (see Fig.~\ref{fig:gravityExpDesign}a for variable definitions). Assuming a ribbon stress of 2 GPa, the suspensions could each support a pair of test masses of $m= 10$ g. If a support of negligible mass could be constructed to provide a lever arm of $r=30$ cm, the devices would theoretically attain $\left(\omega_0/2\pi,Q \right)\approx \left(8 \text{ mHz},6\times 10^{12}\right)$.\footnote{A hypothetical $Q=6\times 10^{12}$ is a factor of 1000 larger than the highest achieved quality factor in thin-film silicon nitride resonators~\cite{cupertino2024centimeter}, highlighting the potential difficulty of the quantum experiment.} For a surface-to-surface separation distance of $200\,\upmu$m ($\approx d-2R$), the modesplitting would be 4 $\upmu$Hz and Eq.~\ref{eq:temperatureRequirement} would be satisfied at 1 mK, depicted in Fig.~\ref{fig:gravityExpDesign}c. If cryogenic cooling of the experiment to 1 mK could be achieved despite challenges such as photothermal heating, a backaction-limited measurement would be possible with an optical lever using a 200 mW laser with 850 nm wavelength and a beam waist of 1 mm (see Appendix~\ref{app:backaction} and Refs.~\cite{pluchar2025imaging,pluchar2025quantum,hao2024back} for more information).

While measuring gravitational entanglement may not be possible with current technology, we note that intermediate path finding experiments would still be capable of exploring new physics. For example, models postulating classical gravity predict excessive noise from the dynamical gravitation between two bodies~\cite{kafri2013noise, kafri2014classical,kryhin2025distinguishable}, and such theories can be probed by near-term experiments~\cite{carney2025quantum}.

\section{Toward tabletop gravity experiments}\label{sec:towardTabletop}
Analyses presented in Section~\ref{sec:GravExp} considered fundamental requirements on pendulum design, optical readout, and cryogenic cooling for an idealized experiment. Various technical obstacles will need to be overcome in practice, particularly for long term ambitions to measure gravitational entanglement. In this section we provide a brief discussion of some of these foreseeable challenges.

\subsection{Non-identical resonance frequencies}
The ideal experiment assumes a pair of identical oscillators, with matched resonance frequencies, such that the mode frequency splitting arises purely from the gravitational interaction. In Appendix~\ref{sec:nonidentical} we analyze the case where the resonance frequencies differ by $\delta\omega$, outlining how gravitational sensitivity deteriorates as $\sigma_{\Delta\omega}\propto \delta \omega$ when the discrepancy exceeds the gravitationally induced modesplitting, i.e. $\delta \omega\gg \Delta \omega$. Therefore, a useful benchmark is to require the resonance frequencies be matched to within the gravitational modesplitting, $\delta\omega \lesssim \Delta \omega$, which for the examples of Section~\ref{sec:GravExp} are $\approx 2\pi \times 1$ $\upmu$Hz.

Matching the resonance frequencies of two torsion pendulums is a demanding test of environmental control and fabrication tolerance. Frequency drift may occur during an experiment due to environmental effects, such as slow variations in the laboratory temperature or in Earth's gravitational field~\cite{adelberger2009torsion}. However, many of these effects will be common to both oscillators. A greater cause for concern is perhaps the construction of identical pendulums. While we anticipate the lithographically defined suspensions to have unparalleled fabrication tolerance, future devices are not expected to achieve such consistency in their test mass distribution or clamping conditions. Possible solutions include
fine tuning the moment of inertia with trim masses~\cite{berg2005laboratory} (difficult in a vacuum chamber) or the stiffness electrostatically~\cite{bai2015improving} or optically~\cite{catano2020high}, or extending the scope of the nanofabrication to encompass more of the test mass system~\cite{bsaibes2025lithographically}.

\subsection{Parasitic electrostatic coupling}
To demonstrate the gravitational coupling between the pendulums it is crucial to eliminate extraneous interactions. Short range gravity experiments have historically been limited by electrostatic interactions, where the use of a conducting shield (omitted from Fig.~\ref{fig:gravityExpDesign}a) to isolate test masses presents a practical limit to their proximity~\cite{adelberger2003tests}. To prevent static charging, the pendulum can be grounded through its suspension. A monofilar Si$_3$N$_4$ suspension can be made conducting through partial metallization~\cite{yu2012control} and a bifilar suspension could be supplemented by a central, fully metallized ribbon that would contribute minimally to the stiffness and dissipation if its width is significantly smaller than the bifilar separation. 

Although the shield may screen direct coupling between the test masses, its electrostatic interaction with each pendulum can introduce deleterious effects such as noise from patch potentials~\cite{ke2023electrostatic,lee2020new,dong2023coupling,speake2003forces}. The magnitude of such effects depends strongly on surface separation~\cite{behunin2012modeling,turchette2000heating}, typically dominating torsion balance experiments only on length scales significantly less than 1 mm~\cite{Kapner2007,lee2020new,tan2020improvement}. The experiment proposed in Section~\ref{sec:classGravExp} allows for a separation of about 7 mm between the shield and each test mass. However, achieving the ultimate goal of measuring gravitational entanglement will likely require greater control over electrostatic effects due to more stringent constraints on proximity and noise (the example discussed in Sec.~\ref{sec:gravEnt} and denoted ``hypothetical" in Fig.~\ref{fig:gravityExpDesign}c would have a shield-to-test-mass separation of about 100 $\upmu$m).

\subsection{Practical limitations to $Q$}\label{sec:lossDiscussion}
We use a simple dissipation dilution model to propose high-$Q$ devices limited purely by material loss (Appendix~\ref{app:pendulumModels}). Silicon-nitride-based resonators have an established track record of achieving exceptional quality factors through stress-induced dissipation dilution~\cite{ghadimi2018elastic,pratt2023nanoscale,sementilli2022nanomechanical}, and the comparatively modest theoretical dilution factor $\approx 10^2$ of the current prototype is sufficient to measure the two-way gravitational interaction. However, an entanglement experiment will likely require quality factors exceeding $10^{10}$, necessitating detailed investigation into extrinsic loss mechanisms such as imperfect clamping conditions~\cite{quinn1995stress}, eddy current damping~\cite{yan2025first,fu2025study}, or gas damping~\cite{cavalleri2010gas}. 

Surface loss presents another practical limitation to $Q$-enhancement via dissipation dilution in thin suspensions, where we have so far assumed $Q\propto h^{-2}$ asymptotically with large dilution. The intrinsic quality factor of transverse modes in Si$_3$N$_4$ has been shown to decrease in thinner films, with an empirical model $Q_\text{int} \approx 60 \times \left(h/1~\text{nm}\right)$ provided by Ref.~~\cite{villanueva2014evidence} where surface loss dominates at roughly sub-$\upmu$m thicknesses. This model suggests a modified scaling $Q\propto h^{-1}$, and predicts a nearly fourfold decrease in $Q_\text{int}$ for a 30 nm film relative to the value of $Q_\text{int}\approx 10^4$ assumed here. While this should not affect the estimated $Q$ for the device in Fig.~\ref{fig:SiN_TB}, which has a $1.8\,\upmu$m thick suspension, it should be considered when optimally designing for a future entanglement experiment that may require extreme dissipation dilution. If surface loss becomes prohibitive, it is possible to ease the constraints on suspension thickness by simultaneously implementing additional dissipation dilution schemes, e.g. an optical spring~\cite{catano2020high}.

\section{Conclusion}
In summary, we have applied nanofabrication techniques to demonstrate a macroscopic torsion pendulum, where a $1.8\,\upmu$m thick suspension supports a test mass of 87 grams. This proof-of-principle device suggests an avenue toward the development of mechanical resonators that can simultaneously achieve the high coherence, low frequency, and large mass necessary for tabletop measurements of gravitational dynamics, with the additional benefits of low fabrication tolerance and a capacity for rapid prototyping.

This study illuminates the scaling behavior of relevant quantities ($\omega_0,Q,\eta,T_\text{ent}$) with pendulum design parameters. Specifically, the use of thin films for bifilar or high-aspect-ratio suspensions is motivated by a boost in coherence from dissipation dilution. Taken to an extreme, one may envision the use of monoatomic films, such as graphene, or a carbon nanotube bifilar suspension. However, the limits of this method of diluting mechanical loss are not known  at this time, requiring further investigation of practical barriers---like those discussed in Section~\ref{sec:lossDiscussion}---to achieving ultra high $Q$. Despite this uncertainty, torsion pendulums with thin-film Si$_3$N$_4$ suspensions provide a test bed for developing the next generation of gravity apparatuses, working toward the ultimate goal of measuring gravitationally induced entanglement and probing theories of quantum gravity.

\section*{Acknowledgments}
We thank Will Terrano and Stephan Schlamminger for useful discussions, John Lawall and Gordon Shaw for feedback on the manuscript, and Raphael Rose for help with a literature search. This work is supported by the Heising-Simons Foundation  through Grant 2023-4467 and an RII UArizona National Labs Partnerships Grant. DJW acknowledges additional support from NSF through award no. 2239735.

\appendix

\section{Modeling torsion pendulum mechanics} \label{app:pendulumModels}
 
The pendulums depicted in Fig.~\ref{fig:gravityExpDesign}a have spherical test masses $m=4\pi \rho R^3/3$, with density $\rho$ and radius $R$, each displaced from the torsion axis by a distance $r$. The total mass and moment of inertia are 
\begin{equation}
    \begin{aligned}
        I &= I_\text{support} + 2m  \left( r^2 + \frac{2}{5} R^2 \right) \approx 2 m r^2\\
        M &= m_\text{support} + 2 m \approx 2 m 
    \end{aligned}
\end{equation}
Expressions provided throughout the main text and appendices often include the approximations above, which assume the test masses dominate the mass of the system and the lever arm $r$ greatly exceeds the radius $R$. 

The torsion constant of the suspension contains contributions from shear $\kappa_{E}$\cite{quinn1997novel} and gravitational $\kappa_g$ stiffness~\cite{pratt2023intersection}
\begin{equation}
    \begin{aligned}
        \kappa_E &= n \frac{E {h}^3 w}{6 l} \\
        \kappa_g &=\frac{Mg}{12 l}\left(3 s^2 \left(n-1\right) + w^2\right).
    \end{aligned}
\end{equation}
Where, $n\in \left\{1,2\right\}$ is the number of tethers composing the suspension; we consider both bifilar and monofilar suspensions. The suspensions have width $w$, length $l$, and thickness $h$. Bifilar suspensions have two tethers, with center-to-center separation $s$. Following Ref.~\cite{pratt2023nanoscale}, we assume the elastic modulus to be $E=250$ GPa. The acceleration due to gravity is $g=9.8$ m$\cdot$s$^{-2}$.

The mechanical quality factor $Q$ is enhanced relative to the intrinsic material loss $Q_\text{int}$ due to dissipation dilution as~\cite{pratt2023nanoscale}
\begin{equation}
	Q = Q_\text{int} \left(1 + \frac{\kappa_g}{\kappa_{E}}\right).
\end{equation}
The mass produces a stress $\sigma=Mg/\left(whn\right)$ in the suspension. The maximum mass load is constrained by the stress it produces in the Si$_3$N$_4$ suspension, limiting the size of the masses. For a prescribed maximum stress $\sigma$, the radius of the test masses is limited to 
\begin{equation}
    R\lesssim\left( \frac{3}{8\pi} \frac{ w h \sigma}{\rho g}n \right)^{1/3}.
\end{equation}

\begin{figure*}[t]
    \centering    \includegraphics[width=1.98\columnwidth,trim= 0in 0in 0in 0in]{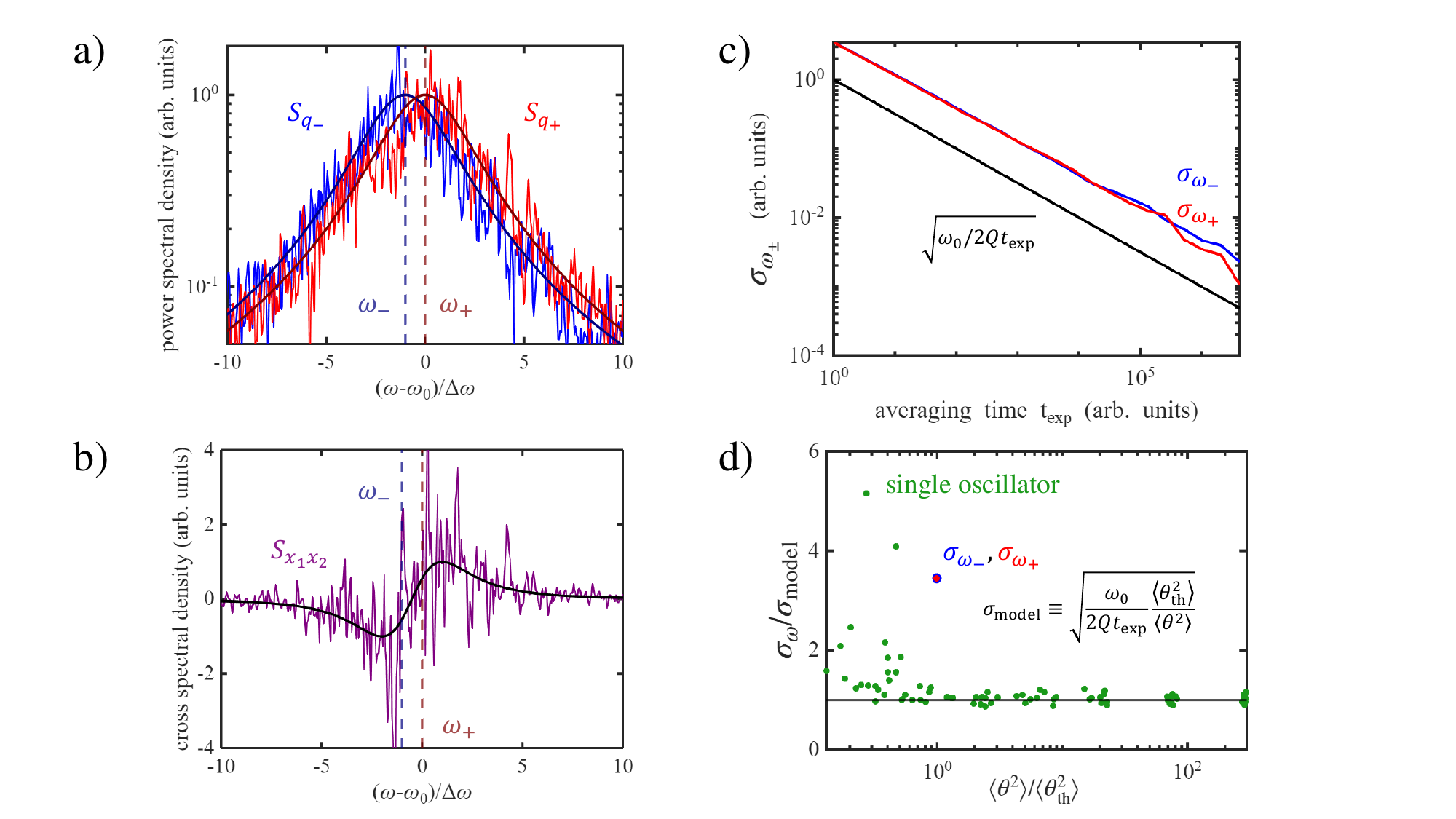}
    \caption{Numerical simulation of coupled oscillators with thermal noise. Simulation results are compared to the analytical models presented in Appendix~\ref{app:classicalModels} for \textbf{a)} the modal PSDs, \textbf{b)} the oscillator position CSD, and \textbf{c)} the individual mode frequency resolution. \textbf{d)} Simulations of a single harmonic oscillator with different starting amplitude show divergence from the frequency resolution model---as occurs in c)---for oscillation amplitudes at or below the thermal motion $\left<\theta_\text{th}^2\right>=k_\text{B}T/I\omega_0^2$.}
    \label{fig:simulation}
\end{figure*}

\section{Model for classical experiment} \label{app:classicalModels}
Here we present a pedagogical overview of the dynamics and observables of an experiment to measure the classical gravitational interaction between two torsion pendulums with resonance frequency $\omega_0$ and moment of inertia $I$, each equipped with spherical test masses $m$, as depicted by Fig.~\ref{fig:gravityExpDesign}(a). Figure~\ref{fig:simulation} provides a visualization of the experimental observables, comparing the analytical models presented here with results from numerical simulations of coupled oscillators. 

\subsection{Gravity-induced mode frequency splitting}\label{sec:modesplitting}
If $\theta_1,\theta_2$ are the coordinates describing the angular displacement of each torsion pendulum, the linear displacement of each mass is $x_i\approx r \theta_i$. Given a separation $d$ between the test mass centers, the gravitational potential energy is
\begin{equation}
    U_G \approx - G m^2\left(\frac{1}{d+x_2-x_1}+\frac{1}{d+x_1-x_2}\right)
\end{equation}
accounting for gravitational coupling only between adjacent masses. Assuming $ \left| x_2- x_1\right|/d \ll 1 $ and expanding to second-order yields
\begin{equation}
    U_G \approx - 2G \frac{m^2}{d} \left( 1 + \frac{r^2}{d^2}( \theta_2 -  \theta_1)^2 \right)
\end{equation}
The system Hamiltonian is 
\begin{equation}
    \begin{aligned}
        H&=\frac{1}{2}I \left(\omega_0^2 \left(\theta_1^2 + \theta_2^2 \right) + \dot{\theta}_1^2 + \dot{\theta}_2^2\right) + U_G
    \end{aligned}    
\end{equation}
yielding coupled equations of motion   
\begin{align*}
    I \ddot{\theta}_1 &= -\kappa_G \left(\theta_2 - \theta_1 \right) - I \omega_0^2  \theta_1 \\
    I \ddot{\theta}_2 &= -\kappa_G \left(\theta_1 - \theta_2 \right) - I \omega_0^2  \theta_2 \\
\end{align*}
where $\kappa_G\equiv 4 G m^2 r^2 / d^3$.

By transforming to the normal mode coordinates $q_\pm=\left(x_2 \pm x_1 \right)/\sqrt{2}$, the equations of motions can be decoupled: $\ddot{q}_\pm = - \omega_\pm^2 q_\pm$. The breathing mode $(-)$ and center-of-mass mode $(+)$ frequencies are  
\begin{equation}
    \begin{aligned}
        \omega_+ &= \omega_0  \\
        \omega_- &= \sqrt{\omega_0^2 - 2\kappa_G/I}
    \end{aligned}
\end{equation}
The mode frequency splitting $\Delta \omega \equiv \omega_+ - \omega_-$ is
\begin{equation} \label{eq:app:modesplitting}
    \Delta \omega \approx \frac{\kappa_G}{I\omega_0} = 2\frac{Gm}{\omega_0 d^3} \left( \frac{2mr^2}{I}\right).
\end{equation}
Assuming spherical masses $m=4\pi\rho R^3 /3$, with $I \approx 2 mr^2$, the center of mass separation must exceed the diameter $d>2 R$, yielding an upper bound on the frequency splitting~\cite{kafri2014classical} 
\begin{equation}
    \Delta \omega< \frac{\pi G\rho}{3 \omega_0}
\end{equation}
While this limitation applies to spherical test masses~\cite{krisnanda2020observable}, more complicated geometries can potentially increase this bound by almost an order of magnitude~\cite{tang2025optimal}.

\subsection{Oscillator and normal mode spectra}
Including dissipation $\gamma=\omega_0/Q$ and uncorrelated thermal torque noise $\left(\left<\tau_1^\text{th} \tau_2^\text{th}\right>=0\right)$ with power spectral density (PSD)
\begin{equation} 
    S_{\tau_i}^\text{th}=S_{\tau}^\text{th} = 4 k_\text{B} T \gamma I.
\end{equation}
into the individual oscillators, the equations of motion become
\begin{equation}
    \ddot{q}_\pm = -\omega_\pm^2 q_\pm - \gamma \dot{q}_\pm + a_\pm^\text{th} 
\end{equation}
Each mode experiences an effective thermal acceleration $a_\pm^\text{th}\equiv \left(\tau_2^\text{th} \pm \tau_1^\text{th}\right) r / \left(\sqrt{2} I \right)$ with equal PSDs
\begin{equation} \label{eq:Sq}
    S_{a_\pm}^\text{th}=S_a^\text{th}= 4 k_\text{B}T\gamma r^2 /I.
\end{equation} 

such that the PSD of each mode is 
\begin{equation}
    S_{q_\pm}= \left|\chi_\pm (\omega)\right|^2 S_a^\text{th}
\end{equation}
where
\begin{equation}
    \chi_\pm(\omega) \equiv \left({\omega_\pm}^2 - \omega^2 - i\gamma \omega \right)^{-1}.
\end{equation}
Although the oscillators are driven by uncorrelated thermal force noise, the gravitational interaction produces a non-zero cross-spectral density $S_{x_1 x_2}$ between them. It can be shown that the PSD $S_{x_i}$ and cross-spectral density $S_{x_1 x_2}$ of the oscillator positions are
\begin{equation}\label{eq:oscillatorPSDCSD}
    \begin{aligned}
        S_{x_i}&=\frac{1}{2} \left( \left|\chi_+\right|^2 + \left|\chi_-\right|^2 \right) S_a^\text{th}\\
        S_{x_1x_2}&=\frac{1}{2} \left( \left|\chi_+\right|^2 - \left|\chi_-\right|^2 \right) S_a^\text{th}
    \end{aligned}
\end{equation}

\subsection{Measuring $\Delta \omega$} \label{sec:sensitivityNoise}
Measurement of the resonance frequency of a thermally driven harmonic oscillator over time $t_\text{exp}$ is subject to uncertainty $\sigma_{\omega_0} \approx \sqrt{\omega_0/\left(2 Q t_\text{exp}\right)}$~\footnote{The full model for $\sigma_{\omega_0}$~\cite{sadeghi2020frequency} accounts for improved performance when the coherent oscillation from a large starting amplitude or a coherent driving force exceeds the thermal motion (see Fig.~\ref{fig:simulation}d). Figure~\ref{fig:simulation}d shows a deviation from the model near thermal equilibrium, where the oscillation amplitude can fall to nearly zero for brief periods of time, resulting in anomalously large frequency estimation errors. This problem can be circumvented by imparting small oscillations above the thermal motion or removing such errors from the measurement record.}. The uncertainty in the normal mode splitting is subject to uncertainty from both modes, $\sigma_{\Delta\omega}^2=\sigma_{\omega_+}^2 + \sigma_{\omega_-}^2\approx 2 \sigma_{\omega_0}^2$, such that 
\begin{equation}	\label{eq:ADthermal}
    \sigma_{\Delta \omega} = \sqrt{\frac{\omega_0}{Q t_\text{exp}}} 
\end{equation}
This model assumes that additive measurement noise is negligible over the measurement bandwidth $t_\text{exp}^{-1}$ about resonance. Assuming white displacement noise $S_\theta^\text{imp}$ in the measurement of each oscillator position, the thermal bandwidth can be shown to be 
\begin{equation}
    \delta \omega_\pm^\text{th} \approx \frac{1}{\omega_0} \sqrt{\frac{S_a^\text{th}}{S_q^\text{imp}}} = \frac{1}{I \omega_0} \sqrt{\frac{S_\tau^\text{th}}{S_\theta^\text{imp}}}
\end{equation}
such that the required measurement noise sensitivity is
\begin{equation}
    S_\theta^\text{imp}\lesssim \frac{S_\tau^\text{th}}{(2\pi)^2 I^2 \omega_0^2}t_\text{exp}^2.
\end{equation}
The performance of a sensor can then be quantified by the measurement time needed to achieve a prescribed signal-to-noise (SNR). For example, to achieve $\Delta \omega / \sigma_{\Delta \omega} = Z$ (or roughly $Z \sigma$-detection) the required measurement time using Eqs.~\ref{eq:app:modesplitting} and~\ref{eq:ADthermal} is\footnote{For small surface separations between the test masses, the center-to-center separation is proportional to the test mass radius, $d\propto R$, such that $d^6\propto m^2$. Therefore, mass $m$ cancels in this expression and the pendulum design parameters are fully accounted for by the figure of merit $\eta$. }
\begin{equation}
    t_\text{exp}=Z^2\frac{\omega_0}{ Q \Delta \omega^2}\approx Z^2\frac{d^6}{4 G^2 m^2} \eta^{-1}
\end{equation}
From this expression we define a figure of merit for comparing torsion pendulum designs, $\eta\equiv Q/\omega_0^3$, which is to be maximized for the best performance.

\subsection{Non-identical oscillators}\label{sec:nonidentical}
The normal modes of the composite oscillator system defined above assume identical oscillators. In practice, the oscillators may not be identical due to fabrication imperfections or the differential influence of environmental effects. The mode frequency splitting is particularly sensitive to discrepancies in the oscillator resonance frequencies, which we model in this section.

We start by introducing a small frequency difference $\delta \omega \ll \omega_0$ between the two oscillators
\begin{equation}
    \begin{aligned}
        \omega_1' &= \omega_0 + \delta \omega/2 \\
        \omega_2' & = \omega_0  - \delta \omega/2
    \end{aligned}
\end{equation}
The analysis of Section~\ref{sec:modesplitting} can be emulated with a redefinition of the modal coordinates
\begin{equation}
    q'_\pm \approx \frac{1}{\sqrt{2}} \left(x_2 +\left(\frac{\delta \omega}{\Delta \omega}\pm\sqrt{1+\frac{\delta \omega^2}{\Delta \omega^2}}\right) x_1\right)
\end{equation}
resulting in decoupled equations of motion with mode frequencies
\begin{equation}
        \omega_\pm' \approx \omega_0  + \frac{\Delta \omega}{2} \left(-1 \pm \sqrt{1+\frac{\delta \omega^2}{\Delta \omega^2}}\right)
\end{equation}
The new mode frequency splitting $\Delta \omega' = \omega_+' - \omega_-'$ is
\begin{equation}
    \Delta \omega' \approx \sqrt{\Delta \omega^2 + \delta \omega^2} 
\end{equation}
For an oscillator frequency offset less than the gravitational mode splitting, $\delta\omega \ll \Delta \omega$, the mode spacing is simply $\Delta\omega'\approx \Delta \omega$. 

However, for a large oscillator frequency offset, $\delta \omega \gg \Delta \omega$, the sensitivity to the gravitational effect is reduced: $\partial \Delta \omega'/\partial \Delta \omega \approx \Delta \omega/\delta \omega$. In this regime, the resolution of the gravitationally induced frequency splitting is 
\begin{equation}
    \sigma_{\Delta \omega} \!\left[\delta \omega \gg \Delta \omega\right] \approx \left|\frac{\partial \Delta \omega'}{\partial \Delta \omega}\right|^{-1}\!\!\!\! \sigma_{\Delta \omega'} \approx \frac{\delta \omega}{\Delta \omega}\sqrt{\frac{\omega_0}{Q t_\text{exp}}} 
\end{equation}

\section{Measurement backaction} \label{app:backaction}
Optical readout of the oscillator motion produces measurement backaction due to radiation pressure shot noise. Here we parameterize the backaction noise of a readout scheme by a measurement rate $\Gamma_\text{meas}$ as 
\begin{equation}
    S_\tau^\text{BA}=4 \hbar \omega_0 I \Gamma_\text{meas}
\end{equation}
in analogy to thermal torque noise $S_\tau^\text{th}=4 \hbar \omega_0 I \Gamma_\text{th}$ with thermal decoherence rate $\Gamma_\text{th}= \gamma \bar{N}_\text{th} \approx  k_\text{B} T /\left(\hbar Q\right)$. Therefore, the criterion that an experiment be in the backaction-limited regime can be stated as $\Gamma_\text{meas}\gg \Gamma_\text{th}$. 

Details of the optical readout scheme are contained within $\Gamma_\text{meas}$. For an optical lever with unity measurement efficiency, where a laser beam with power $P_\text{L}$, optical frequency $\omega_\text{L}$, and waist $w_\text{L}$ is focused on the pendulum at the torsion axis,
\begin{equation}
    \text{optical lever:} \qquad \Gamma_\text{meas}=\frac{\omega_\text{L}P_{L}}{c^2 I\omega_0} \frac{w_\text{L}^2}{2}
\end{equation}
which we derive from the spatiotemporal backaction analysis of Ref.~\cite{pluchar2025imaging}. An alternative readout scheme may employ an optical cavity with finesse $\mathcal{F}$ to measure the linear displacement $x_i$ of one of the test masses, such that~\cite{aspelmeyer2014cavity}
\begin{equation}
    \text{optical cavity:} \qquad \Gamma_\text{meas}=\frac{\omega_\text{L}P_{L}}{c^2 I\omega_0}\frac{8 r^2 \mathcal{F}^2}{\pi^2}
\end{equation}

\bibliography{references.bib}
\end{document}